\documentclass[12pt]{article}
\usepackage{amsmath}
\usepackage{latexsym}
\usepackage{bbm}

\def\g{\gamma}

\def\th{\theta}

\def\m{\mu}
\def\n{\nu}

\def\t{\tau}

\def\pa{\partial}

\def\and{{\rm and}}

\def\ie{{\it i.e.,} }

\def\IR{\mathbbm R}
\def\IZ{\mathbbm Z}

\begin{document}
\vspace*{-1.0in}
\thispagestyle{empty}
\begin{flushright}
CALT-TH-2014-137
\end{flushright}

\parskip = 6pt
\large
\baselineskip = 20pt

\vspace{1.0in} {\Large
\begin{center}
BPS Soliton Solutions of a D3-brane Action
\end{center}}\vspace{.25in}

\begin{center}
John H. Schwarz\footnote{jhs@theory.caltech.edu}
\\
\emph{Walter Burke Institute for Theoretical Physics\\
California Institute of Technology\\ Pasadena, CA  91125, USA}
\end{center}
\vspace{.25in}

\begin{center}
\textbf{Abstract}
\end{center}
\begin{quotation}
\noindent

The world-volume action of a probe D3-brane in
$AdS_5 \times S^5$ with $N$ units of flux
has the field content, symmetries, and dualities
of the $U(1)$ factor of ${\cal N} =4$ $U(N+1)$
super Yang--Mills theory, spontaneously broken to $U(N) \times U(1)$ by being
on the Coulomb branch, with the
massive fields integrated out. Thus, it might be the exact effective action (a
{\em highly effective action}), or else a useful approximation to it. We construct
an $SL(2,\IZ)$ multiplet of BPS soliton solutions of the D3-brane action and
show that in the $N=1$ case they correspond to the electrically charged states
that have been integrated out as well as magnetic monopoles and dyons.
Their charges are uniformly spread on a spherical surface, {\em a soliton bubble},
which can be interpreted as a phase boundary.
This picture is consistent with previous results in the string theory and
field theory literature.

\end{quotation}

\newpage

\pagenumbering{arabic}



The bosonic part of the world-volume action of a probe D3-brane in the
$AdS_5 \times S^5$ solution of Type IIB superstring theory
was constructed in \cite{Schwarz:2013wra}. The symmetries of the background
are induced as symmetries of the D3-brane action. These symmetries include
the superconformal group $PSU(2,2|4)$ and the duality group $SL(2, \IZ)$,
though the bosonic truncation that was presented
only retains the bosonic subgroup of the
superconformal group. The parameters of the theory are a modular parameter
$\t = \chi + i/g_s$, a positive integer $N$, and a mass scale $v$, which encodes
the radial coordinate of the D3-brane in $AdS_5$.  $\chi$ is the background value of
the RR scalar field, the string coupling constant $g_s$ is the exponential of the background
value of the dilaton field, and $N$ is the number of units of five-form flux
that threads the five-sphere.

The physical field content of the D3-brane action consists of a single
abelian ${\cal N} =4$ supermultiplet (a $U(1)$ gauge field, four Majorana spinors,
and six scalars). The inclusion of the fermions in the action has been worked out
for the case of an $AdS_5 \times S^5$ background geometry in \cite{Metsaev:1998hf}.
Other equivalent formulations may be more convenient for our
purposes. In any case, the fermion dependence is not required
for the construction of the classical solutions that are the main goal of this paper,
though it would be useful for verifying some
of the claims that will be made, as well as various other purposes.

The probe D3-brane action in an
$AdS_5 \times S^5$ background with $N$ units of flux gives an
approximate description of the dynamics that is best for large
$N$, when back-reaction effects become negligible. Nevertheless,
the conjecture of \cite{Schwarz:2013wra} is that this action, with
$N$ set equal to 1, is the {\em exact} effective action for
${\cal N} =4$ super Yang--Mills theory
with gauge group $U(2)$ on the Coulomb branch, with all massive fields
integrated out.\footnote{Related ideas involving
$U(N+K) \to U(N) \times U(K)$ are discussed in \cite{Ferrari:2013aba}.
A possible connection to noncommutative field theory is discussed in \cite{Yang:2014xna}.}
Such an action has been dubbed a {\em highly effective action}
or HEA in \cite{Schwarz:2013wra}.
On the Coulomb branch, the electrically charged fields in the $U(2)$ theory are
massive. Once they are integrated out, one obtains an effective
action in terms of the massless neutral fields that correspond to a $U(1)$
subgroup of $U(2)$ (plus a free $U(1)$ multiplet).
This paper will demonstrate that the probe D3-brane action gives exactly
the expected spectrum of half-BPS soliton configurations. 
If the probe D3-brane action is not the exact HEA, it must be a good enough
approximation or truncation to account for this success.

Type IIB superstring theory has an $SL(2,\IZ)$ symmetry group. AdS/CFT
therefore requires that this should be the duality group of the dual
${\cal N} = 4$ Yang--Mills theory. This is the duality group when the
gauge group is $U(N)$, which therefore should be the correct gauge group for the
SYM theory that is dual to the $AdS_5 \times S^5$ type IIB theory
with $N$ units of flux.\footnote{For an up-to-date account of this issue,
as well as earlier references, see \cite{Belov:2004ht}.}
If it were a subgroup, such as $SU(N)$ or $SU(N)/\IZ_N$,
one would obtain the wrong duality group \cite{Aharony:2013hda}.
In the problem of interest we have a probe brane that is separated from the $N$
branes that are responsible for the background flux. This suggests
that we are dealing with a $U(N+1)$ theory
on the Coulomb branch for which the gauge group is spontaneously broken to $U(N) \times U(1)$.
The formulas we will discuss pertain to the $U(1)$ factor. Thus, we only utilize
an abelian gauge field, which results in vast simplifications.
As we have said, the special case $N=1$ is the one that we understand best,
but we hope that our approach is applicable for all $N$.

The modular parameter $\tau$, expressed in terms of standard gauge theory parameters, is
\begin{equation} \label{tau}
\t = \frac{\th}{2\pi} + i\frac{4\pi}{g^2}.
\end{equation}
In the $U(N)$ theory the  $SL(2,\IZ)$  duality group is generated by
$\t \to -1/\t$ and $\t \to \t +1$. The $\t \to -1/\t$ duality of the D3-brane action was
verified in \cite{Schwarz:2013wra}. The usual correspondence between string theory
and gauge theory parameters is
\begin{equation}
\chi = \frac{\th}{2\pi} \quad  \and \quad g_s = \frac{g^2}{4\pi}.
\end{equation}

The bosonic part of the probe D3-brane action in an $AdS_5 \times S^5$
background (in the static gauge) derived in \cite{Schwarz:2013wra} is
\begin{equation}\label{Naction}
S = \frac{1}{2 \pi g_s k^2} \int  \left(\sqrt{-h}
- \sqrt{- \det \left( G_{\m\n} +  k F_{\m\n}\right)} \right)\, d^4 x
+ \frac{\chi}{8\pi} \int F\wedge F,
\end{equation}
where
\begin{equation}
G_{\m\n} = h_{\m\n} + k^2\frac{\pa_\m \phi^I \pa_\n \phi^I}{ \phi^2}.
\end{equation}
and $k = \sqrt{g_s N/\pi}$. Also, $h_{\m\n} = \phi^2 \eta_{\m\n}$,
$ h = \det h_{\m\n}$, and $\eta_{\m\n}$ is the 4d Minkowski metric,
which implies that $\sqrt{-h} = \phi^4$.
There are six scalar fields $\phi^I$, corresponding to the six dimensions transverse
to the D3-brane, and $\phi^2 = \sum(\phi^I)^2 $. The $\int F \wedge F$ term only
contributes for magnetically charged configurations. Then the dependence on $\chi$
is periodic with period 1.

It is interesting to contrast this action with the corresponding
formula for a probe D3-brane in $\IR^{9,1}$. That case has
an action that looks quite similar. However, there is no five-form flux,
so there is no integer $N$. Also, $h_{\m\n} = \eta_{\m\n}$,
$G_{\m\n} = \eta_{\m\n} + k^2 \pa_\m \phi^I \pa_\n \phi^I$, and $k$ is
the reciprocal of the fundamental string tension. Since $k$ is dimensionful,
the action is not scale invariant or superconformal. Rather, it has
${\cal N} =4$ super-Poincar\'e symmetry, and its massless fermions can be
identified as Goldstinos for four spontaneously broken supersymmetries.
This is in contrast to the $AdS_5  \times S^5$ case, which has $PSU(2,2|4)$
superconformal symmetry, part of which is broken spontaneously by the scalar
field vev.

By considering the weak-field limit of Eq.~(\ref{Naction}),
one finds that the fields $\phi^I $ and the gauge field
$A_\m$ are $\sqrt{2\pi g_s}$ times the canonically normalized fields.
Rewriting Eq.~(\ref{Naction}) in terms of the canonically normalized fields, for
which we use the same symbols, one obtains
\begin{equation}\label{Naction2}
S = \frac{1}{\g^2} \int \phi^4 \left(1 -
\sqrt{- \det M_{\m\n}} \right)\, d^4 x
+ \frac{1}{4} g_s \chi\int F\wedge F,
\end{equation}
where
\begin{equation}
M_{\m\n} = \eta_{\m\n} + \g^2\pa_\m \phi^I \pa_\n \phi^I/ \phi^4
+  \g F_{\m\n}/\phi^2
\end{equation}
and
\begin{equation}
\g  = \sqrt{\frac{N}{2\pi^2}}.
\end{equation}
This parameter appears because $\g^2 = R^4 T_{D3}$, where $R$ is the
radius of $AdS_5$ and $T_{D3}$ is the D3-brane tension.
Note that the parameter $g_s$ has completely dropped out of the
first term in the action.
If the true effective action defined by a gauge theory path integral does not share this
property, then this would demonstrate that the D3-brane action is only an
approximation to the HEA.

In the usual formulation with $W$ fields, the Feynman diagrams
of the $U(2)$ theory on the Coulomb branch that contain only massive propagators and
massless external states contain coupling constants in both numerators and denominators.
The vertices give powers of the coupling constant in the
numerator and powers of the mass $M_W = gv$ appear in denominators. So cancellations
of the $g$ dependence are possible. It would be worthwhile to explore this carefully.
The absence of infrared and ultraviolet divergences (when fermions are included)
should make the argument quite clean. The absence of coupling-constant dependence puts
the D3-brane action candidate for the HEA of ${\cal N} =4$ super Yang--Mills theory in 4d
on a comparable footing to the M5-brane action candidate for the HEA of
the (2,0) theory in 6d, which has no adjustable couplings in the first place.
The (2,0) theory is discussed briefly in the appendix.

Once we set $N=1$, the first term in the action no longer contains any parameters.
If we choose to retain the $N$ dependence, a further rescaling of fields by $\sqrt{N}$
brings out all the $N$ dependence of the brane action as an overall factor of $N$.
Thus $S(N) = N S_H$, where $S_H = S(1)$ is given by the formula above with $N=1$.
This shows that the loop expansion of $S(N)$ is a $1/N$ expansion.\footnote{The
analogous loop-expansion parameter for a probe M2-brane action for 3d ABJM theory is $1/\sqrt{kN}$,
and for a probe M5-brane action for the 6d $(2,0)$ theory it is $1/N^2$.} However,
$S_H$, which we claim is relevant to the $U(2)$ theory, has no small parameter.
Even so, we will show that the classical approximation gives the expected half-BPS
solutions.

Let us now turn to the construction of soliton solutions. For the analysis
that follows, it is sufficient to assume that the D3-brane
is at a fixed position on the five-sphere.
Then the five scalars fields that correspond to the $S^5$
factor in the spacetime geometry do not contribute.
If we are interested in spherically symmetrical static solutions, centered at $r=0$,
we may assume that the electric and magnetic fields $\vec E$ and $\vec B$ only
have nonzero radial components, denoted $E$ and $B$, and that all fields ($E$, $B$, and $\phi$)
are functions of the radial coordinate $r$ only. Then
$\det(-G_{\m\n}) = \phi^6 G_{rr}$, where
\begin{equation}
G_{rr} = \phi^2 + \g^2 (\phi'/\phi)^2,
\end{equation}
and
\begin{equation}
-\det (G_{\m\n} + \g F_{\m\n})  = \phi^6 \left(G_{rr}  - \frac{\g^2 E^2}{\phi^2}\right)
\left( 1 + \frac{\g^2 B^2}{\phi^4} \right).
\end{equation}
This results in the Lagrangian density
\begin{equation}\label{gamma}
{\cal L} = \frac{1}{\g^2}  \phi^4\left(1 -
\sqrt{\left( 1 + \frac{\g^2 [(\phi')^2-E^2]}{\phi^4} \right)
\left(1 +  \frac{\g^2 B^2}{\phi^4}\right)} \, \right) +g_s \chi BE.
\end{equation}

The field canonically conjugate to $A_r$ is
\begin{equation} \label{Deqn}
D =  \frac{\pa {\cal L}}{\pa E} = E \sqrt{\frac{1+\g^2 B^2/\phi^4}{1 + \g^2 [(\phi')^2 - E^2]/\phi^4}} +  g_s \chi B,
\end{equation}
and the energy density ${\cal H}  = DE - {\cal L}$ is
\begin{equation}
{\cal H}  =  \frac{1}{\g^2} \phi^4\left((1 + \g^2 (\phi')^2/\phi^4)
\sqrt {\frac{1 + \g^2 B^2 /\phi^4} {1 + \g^2 [(\phi')^2 -E^2]/\phi^4}} -1 \right).
\end{equation}
Solving Eq.~(\ref{Deqn}) for $E$ gives
\begin{equation} \label{Eeqn}
E =  \tilde D \sqrt{\frac{1+\g^2 (\phi')^2/\phi^4}{1 + \g^2 [B^2 + \tilde D^2]/\phi^4}},
\end{equation}
where
\begin{equation}
\tilde D = D - g_s \chi B.
\end{equation}
Eliminating $E$ in favor of $\tilde D$ in the energy density then gives
\begin{equation} \label{ham}
{\cal H}  = \frac{1}{\g^2} \phi^4\left(\sqrt {(1 + \g^2 (\phi')^2/\phi^4)
(1 + \g^2 (B^2 + \tilde D^2)/\phi^4)} -1 \right).
\end{equation}

The equation of motion for $A_0$, the time component of the gauge field,
in regions with vanishing electric-charge density, is
\begin{equation}
\frac{\pa}{\pa r} (r^2 D) =0.
\end{equation}
Similarly, $B$ satisfies the same equation in regions of vanishing magnetic-charge
density. For a soliton centered at $r=0$, with $p$ units of electric charge $g$ and $q$ units
of magnetic charge $g_m$, where $g_m = 4\pi/g$,
we have
\begin{equation}\label{D2B2}
D =  \frac{p g}{4\pi r^2} \quad
\and \quad B = - \frac{q g_m}{4\pi r^2}.
\end{equation}
The minus sign, which is introduced for later convenience, is a matter of convention.
Thus,
\begin{equation}
B^2 + \tilde D^2 = Q^2/r^4,
\end{equation}
where
\begin{equation} \label{Qeqn}
Q = \frac{g}{4\pi} |p + q\t|,
\end{equation}
and $\t$ is defined in Eq.~(\ref{tau}).
An unusual feature of this analysis is that we have introduced
the unit of electric charge $g$ and the angle $\th$ as external data, since
only the combination $g^2 \th$ is present in the action (\ref{Naction2}).
We will return to this issue later.

Let us now turn to the problem of finding BPS solutions. Specifically, we wish
to find functions $\phi(r)$ that give BPS extrema of $H = 4\pi \int {\cal H} r^2 dr$, where
\begin{equation} \label{BPSmass}
{\cal H}  = \frac{1}{\g^2}\left(\sqrt {(\phi^4 + \g^2 (\phi')^2)
(\phi^4 + \g^2 Q^2/r^4)} -\phi^4 \right),
\end{equation}
subject to the requirement that $\phi \to v$ as $r \to \infty$, which specifies the radial position
of the unperturbed D3-brane in $AdS_5$. This also makes contact with the parameter $v$ in the
$U(2)$ theory on the Coulomb branch in the standard notation.
A plausible guess is that the BPS condition requires that the two factors inside the
square root in Eq.~(\ref{BPSmass}) should be equal, \ie
\begin{equation} \label{BPS}
(\phi')^2 =  Q^2/r^4,
\end{equation}
which implies that ${\cal H} = (\phi')^2$. The proof that this is BPS goes as
follows.\footnote{This line of reasoning was suggested by Jaemo Park.} One first rewrites
Eq.~(\ref{BPSmass}) in the form
\begin{equation}
(\g^2 {\cal H} + \phi^4)^2 = (\g^2 X |\phi'| + \phi^4)^2 + \g^2 \phi^4 (X - |\phi'|)^2,
\end{equation}
where $X = Q/r^2$. Thus,
\begin{equation}
 (\g^2 {\cal H} + \phi^4)^2 \geq (\g^2 X |\phi'| + \phi^4)^2,
\end{equation}
which implies ${\cal H} \geq X|\phi'|$. Equality requires $|\phi'| =X$, which then gives
${\cal H} = X^2 = (\phi')^2$.
It is also worth pointing out that Eq.~(\ref{Eeqn}) implies that $E = \tilde D$
for BPS configurations.
For a given pair of integers $p$ and $q$, there are two BPS solutions of Eq.~(\ref{BPS})
\begin{equation}
\phi_{\pm} = v \pm Q/r .
\end{equation}

In the case of a probe D3-brane embedded in $\IR^{9,1}$,
analyzed in \cite{Callan:1997kz}, the BPS conditions and solutions are similar
to what we are finding. However, there are important
differences. In the flat-space case, $\phi$ corresponds to a Cartesian coordinate
for one of the directions transverse to the brane, and thus it can range from $-\infty$
to $+\infty$. The two solutions $\phi_{\pm}$ in that problem are physically equivalent.
The distinction between them is which side of the D3-brane
has a funnel-shaped protrusion. As noted
in \cite{Callan:1997kz}, these protrusions can be interpreted as $(p,q)$ strings that terminate
on the brane. Given that the strings extend to infinity, it is not surprising that
the mass of these configurations is given by a divergent expression proportional to
$\int dr/r^2$. In fact, one can derive the tension of a $(p,q)$ string by cutting off
the integral and examining the dependence on the cutoff.

Returning to the case of $AdS_5 \times S^5$, the six dimensions transverse to the brane
have been described in spherical coordinates. The $AdS_5$ radial coordinate corresponds
to $\phi$, which ranges from $0$ to $+\infty$. The $\phi_+$ solution is similar to the
flat-space solutions. As $r\to  0$, $\phi \to +\infty$, which corresponds to the boundary
of $AdS_5$. As in flat space, this gives a divergent mass proportional to
$\int dr/r^2$. However, the physical interpretation
in terms of a $(p,q)$ string does not work well, because the cross section of the funnel is
shrinking too quickly. In any case, this is not the solution we are after.

The other BPS solution, $\phi_-(r)$, behaves differently from the flat-space BPS solutions.
The crucial fact is that $\phi =0 $ corresponds
to the {\em horizon of the Poincar\'e patch} of $AdS_5$. More precisely, as explained by
H. Ooguri, it is where the horizon
of the Poincar\'e patch intersects the boundary of global AdS.
Thus there is no continuation beyond this point.
This means that the integral should be cut off at
\begin{equation}
r_0 = \frac{Q}{v}.
\end{equation}
Using Eq.~(\ref{Qeqn}), it follows that the mass of this BPS soliton is
\begin{equation}
M = 4\pi \int_{r_0}^{\infty} {\cal H} r^2 dr
= 4\pi \int_{r_0}^{\infty} (\phi_-')^2 r^2 dr
= \frac{4\pi Q^2}{r_0} = v g |p+q\t|.
\end{equation}
This is exactly the expected answer! There is a stable soliton for each nonzero pair of
coprime integers $(p,q)$. These form an irreducible multiplet of the $SL(2,\IZ)$ duality
group.
In particular, the $(p,q) = (\pm 1,0)$ solitons
(the $W^{\pm}$ bosons) have mass $vg$ and the $(p,q) = (0,\pm 1)$ solitons (the
magnetic monopoles) have mass $4\pi v/g$ (for $\th =0$). (See section 3.3 of \cite{Shifman:2009zz}
for a pedagogical review of 't Hooft--Polyakov monopoles in the Georgi--Glashow model.)
The soliton charges are spread uniformly on the sphere $r = r_0$,
which we propose to call a {\em soliton bubble}.

The soliton masses saturate the BPS bound given by the central charges in the supersymmetry
algebra. While this is a convincing argument that these solutions are half-BPS, it would
be nice to demonstrate explicitly that they preserve half of the supersymmetries.
We hope to do that in the future when we incorporate the fermi fields in the theory.
We have treated the brane action in the classical approximation, but the results
are expected to remain valid in the quantum theory, since properties of half-BPS
states in theories with this much supersymmetry should be robust.

Spherical shells of charge, like the soliton bubbles found here, have appeared
previously. Gauntlett {\it et al.} \cite{Gauntlett:1999xz}
studied a probe D3-brane in an asymptotically flat black D3-brane supergravity
background. This is a closely related problem, since this geometry has $AdS_5 \times S^5$
as its near-horizon limit \cite{Maldacena:1997re}. They identified half-BPS solutions,
like those found here, ``in which a point charge is replaced by a perfectly conducting
spherical shell.'' They also discussed quarter-BPS configurations that are related to string
junctions.

Soliton bubbles have some striking analogies with supersymmetric
black holes. This is surprising, inasmuch as the 4d field theory is nongravitational
and resides on a flat spacetime. For one thing,
as one approaches the soliton bubble from the outside,
the scalar field approaches $\phi(r_0) =0$, the value
of the field for which the Coulomb-branch description of the theory breaks down,
irrespective of its value at spatial infinity.
This is reminiscent of the attractor mechanism for ${\cal N} =2$ extremal black
holes \cite{Ferrara:1995ih}. In fact, using attractor flow equations,
Denef \cite{Denef:1998sv}, \cite{Denef:2000nb}
(see also \cite{Denef:2001xn}, \cite{Denef:2007vg})
found similar structures to our soliton
bubbles in the context of supergravity solutions in which a D3-brane wraps a cycle
of a Calabi--Yau manifold that vanishes at a conifold point, where the central
charge modulus is zero. He calls the resulting BPS solitons ``empty holes.''
It seems reasonable to suppose that there are empty holes, or bubbles of nothing,
in the gravitational context that he analyzed.
However, in a flat-space matter theory, which is what we are considering, that seems
implausible. Therefore, we will make an alternative proposal shortly.

For any choice of $(p,q)$, the mass of the probe D3-brane soliton is proportional
to the radius of the soliton bubble with a universal coefficient
\begin{equation}
M = 4\pi v^2 r_0 .
\end{equation}
For comparison, the mass of an extremal Reissner--Nordstrom
black hole in four dimensions is related to the radius of its horizon
by $M=r_0/G$, where $G$ is Newton's constant. Thus, $v$ is the analog
of the Planck mass. One also has $Mr_0 \sim Q^2$ in both cases.
According to \cite{Swingle:2014uza} and references therein,
gravity in 10d should correspond to quantum entanglement of the dual 4d CFT.
Together with the black-hole analogy, this suggests that, up to a numerical
coefficient, $Q^2$ should measure the entanglement
entropy between the inside and outside of the soliton bubble.

Since the BPS solitons of the D3-brane probe's action are solutions of a nongravitational theory in
Minkowski spacetime, they are not black holes or bubbles of nothing in any
conventional sense. So what is happening? An interpretation (suggested by Abhijit Gadde)
is that the gauge theory is in the conformal phase inside the sphere. More precisely,
it is in the ground state of the conformal phase, since the interior should
not contribute to the mass of the soliton. This implies that the soliton bubble
is a {\em phase boundary}. (Even though soliton bubbles have appeared previously,
their interpretation as phase boundaries appears to be new.) In order to describe the
theory in the conformal phase, which appears in the interior of soliton bubbles,
the charge $g$ and the angle $\th$ need to be specified.
This would explain where these parameters comes from. 

In view of the black-hole analogy,
discussed in the preceding paragraphs, one is now tempted to ask
whether the horizon of an extremal Reissner-Nordstrom black hole
could have an interpretation as a phase boundary. This may sound preposterous,
yet something like that has been proposed by Dvali and Gomez \cite{Dvali:2012en}.
It is not obvious how to relate their discussion to ours, but it may be
worthwhile to try to do so.

Similar phenomena have appeared also in string theoretic studies with ${\cal N} =2 $ supersymmetry,
in addition to those mentioned previously \cite{Denef:1998sv}, \cite{Denef:2000nb}.
In \cite{deMelloKoch:1999ui} the authors utilized the DBI action of a D3-brane probe in F theory
and argued that it includes non-holomorphic higher-derivative corrections to the Seiberg--Witten
effective action. They constructed a monopole solution containing a soliton bubble that
coincides with a 7-brane. Monodromies of the 7-brane correspond to dualities of the gauge theory.

The formation of soliton bubbles may also be related to the {\em enhan\c{c}on} mechanism
in \cite{Johnson:1999qt}, which circumvents the appearance of a class of naked singularities,
known as {\em repulsons}. The specific context that was considered
is related to large-$N$ Seiberg--Witten theory.

In \cite{Popescu:2001rf}, Popescu and Shapere studied the low-energy effective action
of ${\cal N} =2 $ $SU(2)$ gauge theory with no matter in the Coulomb phase \cite{Seiberg:1994rs}.
They obtained the monopole and dyon hypermultiplets as BPS solutions
exhibiting a spherical shell of charge. The radius of the shell is given by $Z(r_0) =0$, where
the ``local central charge'' is $Z(r) = n_e a(r) + n_m a_D (r)$. Vanishing local central charge
is required for nonvanishing charge.
Since $a_D(r_0)/a(r_0)$ is real, this shell is on the wall of marginal stability,
and it is a phase boundary. As in our ${\cal N} =4$ example, a different action is required to
describe the interior phase. It is noteworthy that the basic picture of soliton bubbles
associated to phase boundaries arises in a nonconformal theory. The paper \cite{Popescu:2001rf},
as well as the ones discussed previously, did not examine whether the
charged vector multiplets of ${\cal N} =2$ theories can be obtained
as solitons in a similar manner. In the ${\cal N}=4$ case this was guaranteed, because they
are related to the monopoles by the duality group. In the ${\cal N} =2$ theories that is not
the case.

There is some interesting related evidence for
soliton bubbles \cite{Bolognesi:2005rk}--\cite{Taubes:2013xpa} in a nonsupersymmetric
field theory context.\footnote{The author is grateful to David Tong and Nick Manton
for bringing this work to his attention.} By considering multi-monopole solutions of large
magnetic charge in the 4d $SU(2)$ gauge theory with adjoint scalars on the
Coulomb branch, Bolognesi deduced the existence of ``magnetic bags'' with properties
that are very close to those of soliton bubbles. He also pointed out the
analogy to black holes \cite{Bolognesi:2010xt}. The reason
that this phenomenon was not discovered earlier by considering a single monopole is that the
nonabelian gauge theory solution gives $\phi^a (\vec x) = \frac{x^a}{r} \phi(r)$ with
\begin{equation}
\phi(r) = v ({\rm coth} y - 1/y),
\end{equation}
where $y = gvr$ (for $\th =0$). This differs from $\phi_-(r) = v(1 - 1/y)$ by a series of terms
of the form $\exp(-2n M_W r)$, where $M_W =gv$ and $n$ is a positive integer. Yet,
$\phi(r)$ is strictly positive for $r>0$, and $\phi^a (\vec x)$
is nonsingular at the origin. At least for ${\cal N} =4$, the effect of integrating out the fields
of mass $M_W$ should be to cancel the exponential terms for $y>1$ and to give $\phi =0$ for $y<1$. After all, the $U(1)$ HEA is supposed to incorporate all of those contributions. This
point is also made in an ${\cal N} =2$ context in \cite{deMelloKoch:1999ui}.

It is easy to guess (and to verify) the generalization of our static one-soliton solution
to the case of $n$ solitons of equal charge.
Since the forces between them should cancel when they are at rest, their centers can be
at arbitrary spatial positions $\vec x_k$, $ k = 1,2,\ldots,n$. The solution is then given by
\begin{equation} \label{multi}
\phi (\vec x) = v - Q \sum_{k=1}^n \frac{1}{|\vec x - \vec x_k|}.
\end{equation}
The surfaces of the bubbles are given by $\phi(\vec x) =0$. Clearly, they are no longer
spherical when there is more than one center. It is also easy to visualize how the bubbles merge
or split apart as their positions are varied. The fields $\vec D$ and $\vec B$ are then
proportional to $ \vec\pa \phi$, with coefficients determined by the charges. Also, using the
BPS condition, $\vec E = \vec D - \frac{g^2\th}{8\pi^2} \vec B$. The charge distributions
on the bubbles can then be deduced from the discontinuity in the electric and magnetic fields
by standard methods.
These multi-soliton formulas are a lot simpler than the usual multi-monopole
ones! One reason is that abelian fields are easier to describe than nonabelian ones.
Another is that the quantum effects of the massive fields are incorporated in the formula,
and they cancel exponential terms that would appear otherwise.

Since the classical analysis described above works for any value of $N$, it is tempting
to take the parameter $N$ seriously. The natural context for this is the
Coulomb branch in which the gauge symmetry is broken according to $U(N+1) \to U(1)  \times
U(N)$. The action for the $U(1)$ term should then be $NS_H$. For those solitons
that are singlets of $U(N)$ (or the appropriate dual $U(N)$), the D3-brane action
would be sufficient to describe the
exterior of soliton bubbles, where $\phi >0$, whereas the $U(N+1)$ action,
with the vacuum in the conformal phase, would describe the interiors of soliton bubbles.
However, there should also be solitons that are not singlets of $U(N)$
(or a dual $U(N)$). They would source both the $U(1)$ fields and the
$U(N)$ fields. Just as the $W$ fields, belong to the fundamental and anti-fundamental
representations of the ``electric'' $U(N)$, the basic monopoles are known to
belong to the fundamental and anti-fundamental representations of
the magnetic dual $U(N)$. (For a discussion of the monopoles
and related facts see \cite{Auzzi:2004if} and references therein.) One might
hope that solitons that share charges of different gauge group factors would be
eliminated due to confinement. While that is not the case, in general, certain classes
of solitons will be confined following further symmetry breaking of the remaining
non-abelian factor \cite{Auzzi:2003fs}. To understand soliton solutions that carry
both $U(1)$ and $U(N)$ charges, it is probably necessary to understand the effective
action involving both $U(1)$ and $U(N)$ fields and their mutual interactions.
At best, the D3-brane action only captures the $U(1)$ truncation of this system.
This should be adequate for constructing solitons that do not source the $U(N)$ fields.

${\cal N} =4$ super Yang--Mills theories with $N>1$ poses numerous challenges,
such as understanding additional classes of solitons with less supersymmetry and wall-crossing
phenomena. These occur when D3-branes are localized at different points on
the five-sphere. The D3-brane action described here is expected to play a role
into the formulation of such theories, but we expect there to be additional terms that are 
much more complicated. It is because of these complexities that we have emphasized the case 
$N=1$ as the one we understand best.

If the $N>1$ case were understood well enough,
we could consider large $N$. Then the theory defined by the D3-brane action would have
a controlled loop expansion. Also, if the D3-brane action is only an approximation to the
HEA, the approximation should improve for large $N$. One lesson we have learned is
that all phases of the theory need to be understood, since soliton solutions can
involve multiple phases. One could envisage bubbles within bubbles, corresponding to
sequential symmetry enhancement. 

In conclusion, the conjecture that the D3-brane action in
$AdS_5 \times S^5$ with one unit of flux
is an HEA for the gauge group $U(2)$ has passed a nontrivial test:
it has the expected half-BPS soliton solutions.
Nevertheless, the probe D3-brane action might be too simple to be the exact HEA,
as we have conjectured. What we have
demonstrated here is that it seems to be adequate for revealing supersymmetry-protected
aspects of the $U(2)$ gauge theory that are otherwise hard to access.
Clearly, there are many other directions to explore in the future, some of which
were mentioned in the preceding paragraphs. The results that are expected
for 6d (2,0) theories are sketched in the appendix. Other interesting
problems include clarifying the precise relationship between probe-brane
actions and HEAs as well as the relationship between BPS solitons and
supersymmetric black holes.

\newpage

The author wishes to acknowledge discussions and correspondence with Frank Ferrari, Abhijit Gadde,
Jerome Gauntlett, Nicholas Hunter-Jones, Elias Kiritsis, Arthur Lipstein, Nick Manton,
Hirosi Ooguri, Jaemo Park, Nati Seiberg, and David Tong.
He is also grateful to Yuji Tachikawa for reading a draft of the manuscript
and making helpful comments. This work was supported in part by DOE Grant \# DE-SC0011632.

\newpage

\centerline{\bf Appendix: BPS solitons of the 6d (2,0) theory}

Reference \cite{Schwarz:2013wra} conjectured that the world-volume action of an
M5-brane in $AdS_7 \times S^4$ is the HEA of the $A_1$ (or $U(2)$) 6d
$(2,0)$ theory. It is not our purpose to explore that theory in detail here. However,
based on what we have learned about the D3-brane theory, it is clear what one should
find. There should be static half-BPS soliton solutions that
describe infinitely-extended strings carrying a self-dual charge. In fact,
up to numerical coefficients, we can even write down the formulas. Let us take the
strings to be oriented along the $x^5$ direction with transverse positions
described by four-vectors $\vec x_k$. (The position on the four-sphere is fixed, as before.)
Since the scalar field $\phi$ has dimension two in this case, let us denote its asymptotic
value for large $|\vec x|$ by $v^2$. Then the analog of Eq.~(\ref{multi}) should be
\begin{equation}
\phi (\vec x) = v^2 - \sum_{k=1}^n \frac{1}{|\vec x - \vec x_k|^2}.
\end{equation}
As before, the locus $\phi(\vec x) =0$ describes soliton bubbles whose interiors should be
in the conformal phase.  Also, the self-dual three-form field should take the form
$H_{0i5} \sim \pa_i \phi$.

If all the centers coincide at $r = |\vec x| =0$, then the formula becomes
\begin{equation}
\phi (r) = v^2 - n/r^2.
\end{equation}
The critical radius at which $\phi$ vanishes is $r_0 = \sqrt{n}/v$ and the tension
is $T = n v^2$. These formulas
can be compared to an extremal black string of charge $n$ in six dimensions. Identifying
the radius $r_0$ with the horizon radius of the black string, and $v$ with the Planck mass
in six dimensions, one again finds that the formulas match, at least up to numerical coefficients. In this case, one can conjecture that $n^{3/2} v$ might correspond to entanglement
entropy per unit length. A curious fact about the (2,0) theory is that there is
an action for an abelian factor in the Coulomb phase even though it is generally believed that
the conformal phase has no Lagrangian description.



\newpage

\begin{thebibliography}{99}

\bibitem{Schwarz:2013wra}
  J.~H.~Schwarz,
  ``Highly effective actions,''
  JHEP {\bf 1401}, 088 (2014)
  [arXiv:1311.0305 [hep-th]].

\bibitem{Metsaev:1998hf}
  R.~R.~Metsaev and A.~A.~Tseytlin,
  ``Supersymmetric D3-brane action in AdS(5) x S**5,''
  Phys.\ Lett.\ B {\bf 436}, 281 (1998)
  [hep-th/9806095].

\bibitem{Ferrari:2013aba}
  F.~Ferrari,
  ``Gauge theories, D-branes and holography,''
  Nucl.\ Phys.\ B {\bf 880}, 247 (2014)
  [arXiv:1310.6788 [hep-th]].

\bibitem{Yang:2014xna}
  H.~S.~Yang,
  ``Highly effective action from large $N$ gauge fields,''
  arXiv:1402.5134 [hep-th].

\bibitem{Belov:2004ht}
  D.~Belov and G.~W.~Moore,
  ``Conformal blocks for AdS(5) singletons,''
  hep-th/0412167.

\bibitem{Aharony:2013hda}
  O.~Aharony, N.~Seiberg and Y.~Tachikawa,
  ``Reading between the lines of four-dimensional gauge theories,''
  JHEP {\bf 1308}, 115 (2013)
  [arXiv:1305.0318 [hep-th]].

\bibitem{Callan:1997kz}
  C.~G.~Callan and J.~M.~Maldacena,
  ``Brane dynamics from the Born-Infeld action,''
  Nucl.\ Phys.\ B {\bf 513}, 198 (1998)
  [hep-th/9708147].

\bibitem{Shifman:2009zz}
  M.~Shifman and A.~Yung, {\it Supersymmetric Solitons},
  Cambridge, UK: Cambridge Univ. Pr. (2009) 259 pp.

\bibitem{Gauntlett:1999xz}
  J.~P.~Gauntlett, C.~Kohl, D.~Mateos, P.~K.~Townsend and M.~Zamaklar,
  ``Finite energy Dirac-Born-Infeld monopoles and string junctions,''
  Phys.\ Rev.\ D {\bf 60}, 045004 (1999)
  [hep-th/9903156].

\bibitem{Maldacena:1997re}
  J.~M.~Maldacena,
  ``The large-$N$ limit of superconformal field theories and supergravity,''
  Adv.\ Theor.\ Math.\ Phys.\  {\bf 2}, 231 (1998)
  [hep-th/9711200].

\bibitem{Ferrara:1995ih}
  S.~Ferrara, R.~Kallosh and A.~Strominger,
  ``N=2 extremal black holes,''
  Phys.\ Rev.\ D {\bf 52}, 5412 (1995)
  [hep-th/9508072].

\bibitem{Denef:1998sv}
  F.~Denef,
  ``Attractors at weak gravity,''
  Nucl.\ Phys.\ B {\bf 547}, 201 (1999)
  [hep-th/9812049].

\bibitem{Denef:2000nb}
  F.~Denef,
  ``Supergravity flows and D-brane stability,''
  JHEP {\bf 0008}, 050 (2000)
  [hep-th/0005049].

\bibitem{Denef:2001xn}
  F.~Denef, B.~R.~Greene and M.~Raugas,
  ``Split attractor flows and the spectrum of BPS D-branes on the quintic,''
  JHEP {\bf 0105}, 012 (2001)
  [hep-th/0101135].

\bibitem{Denef:2007vg}
  F.~Denef and G.~W.~Moore,
  ``Split states, entropy enigmas, holes and halos,''
  JHEP {\bf 1111}, 129 (2011)
  [hep-th/0702146 [HEP-TH]].

\bibitem{Swingle:2014uza}
  B.~Swingle and M.~Van Raamsdonk,
  ``Universality of gravity from entanglement,''
  arXiv:1405.2933 [hep-th].

\bibitem{Dvali:2012en}
  G.~Dvali and C.~Gomez,
  ``Black holes as critical point of quantum phase transition,''
  Eur.\ Phys.\ J.\ C {\bf 74}, 2752 (2014)
  [arXiv:1207.4059 [hep-th]].

\bibitem{deMelloKoch:1999ui}
  R.~de Mello Koch, A.~Paulin-Campbell and J.~P.~Rodrigues,
  ``Monopole dynamics in $N=2$ super Yang-Mills theory from a three-brane probe,''
  Nucl.\ Phys.\ B {\bf 559}, 143 (1999)
  [hep-th/9903207].

\bibitem{Johnson:1999qt}
  C.~V.~Johnson, A.~W.~Peet and J.~Polchinski,
  ``Gauge theory and the excision of repulson singularities,''
  Phys.\ Rev.\ D {\bf 61}, 086001 (2000)
  [hep-th/9911161].

\bibitem{Popescu:2001rf}
  I.~A.~Popescu and A.~D.~Shapere,
  ``BPS equations, BPS states, and central charge of $N=2$ supersymmetric gauge theories,''
  JHEP {\bf 0210}, 033 (2002)
  [hep-th/0102169].

\bibitem{Seiberg:1994rs}
  N.~Seiberg and E.~Witten,
  ``Electric-magnetic duality, monopole condensation, and confinement in $N=2$ supersymmetric Yang-Mills theory,''
  Nucl.\ Phys.\ B {\bf 426}, 19 (1994)
  [Erratum-ibid.\ B {\bf 430}, 485 (1994)]
  [hep-th/9407087].

\bibitem{Bolognesi:2005rk}
  S.~Bolognesi,
  ``Multi-monopoles, magnetic bags, bions and the monopole cosmological problem,''
  Nucl.\ Phys.\ B {\bf 752}, 93 (2006)
  [hep-th/0512133].

\bibitem{Lee:2008ze}
  K.~-M.~Lee and E.~J.~Weinberg,
  ``BPS magnetic monopole bags,''
  Phys.\ Rev.\ D {\bf 79}, 025013 (2009)
  [arXiv:0810.4962 [hep-th]].

\bibitem{Bolognesi:2010xt}
  S.~Bolognesi,
  ``Magnetic bags and black holes,''
  Nucl.\ Phys.\ B {\bf 845}, 324 (2011)
  [arXiv:1005.4642 [hep-th]].

\bibitem{Harland:2011tm}
  D.~Harland,
  ``The Large $N$ limit of the Nahm transform,''
  Commun.\ Math.\ Phys.\  {\bf 311}, 689 (2012)
  [arXiv:1102.3048 [hep-th]].

\bibitem{Manton:2011vm}
  N.~S.~Manton,
  ``Monopole planets and galaxies,''
  Phys.\ Rev.\ D {\bf 85}, 045022 (2012)
  [arXiv:1111.2934 [hep-th]].

\bibitem{Harland:2012cj}
  D.~Harland, S.~Palmer and C.~Saemann,
  ``Magnetic domains,''
  JHEP {\bf 1210}, 167 (2012)
  [arXiv:1204.6685 [hep-th]].

\bibitem{Taubes:2013xpa}
  C.~H.~Taubes,
  ``Magnetic bag like solutions to the SU(2) monopole equations on $R^3$,''
  arXiv:1302.5314 [math-ph].

\bibitem{Auzzi:2004if}
  R.~Auzzi, S.~Bolognesi, J.~Evslin, K.~Konishi and H.~Murayama,
  ``Non-Abelian monopoles,''
  Nucl.\ Phys.\ B {\bf 701}, 207 (2004)
  [hep-th/0405070].

\bibitem{Auzzi:2003fs}
  R.~Auzzi, S.~Bolognesi, J.~Evslin, K.~Konishi and A.~Yung,
  ``Non-Abelian superconductors: vortices and confinement in $N=2$ SQCD,''
  Nucl.\ Phys.\ B {\bf 673}, 187 (2003)
  [hep-th/0307287].

\end{thebibliography}
\end{document}